\documentclass[fleqn,10pt]{wlscirep}
\usepackage[utf8]{inputenc}
\usepackage[T1]{fontenc}
\usepackage[version=3]{mhchem}

\usepackage{pdfpages}

\usepackage{multicol}
\usepackage{array}
\usepackage{hyperref}

\usepackage{xcolor}

\usepackage{pdfpages}

\usepackage{graphicx}
\graphicspath{{figures/}{../figures/}}

\newcommand{\comment}[1]{}

\usepackage{titlesec}

\titleformat{\paragraph}
{\normalfont\normalsize\bfseries}{\theparagraph}{1em}{}
\titlespacing*{\paragraph}
{0pt}{3.25ex plus 1ex minus .2ex}{1.5ex plus .2ex}

\title{High throughput inverse design and Bayesian optimization of functionalities: spin splitting in two-dimensional
compounds}
\author[1,a]{Gabriel M. Nascimento}
\author[1,a]{Elton Ogoshi}
\author[1,2]{Adalberto Fazzio}
\author[1,*]{Carlos Mera Acosta}
\author[1,*]{Gustavo M. Dalpian}
\affil[a]{These authors equally contributed  to this work.}
\affil[1]{Center for Natural and Human Sciences, Federal University of ABC, Santo Andre, SP, Brazil}
\affil[2]{Brazilian Nanotechnology National Laboratory (LNNano), CNPEM, 13083‐970, Campinas, São Paulo, Brazil}

\affil[*]{Corresponding authors: mera.acosta@ufabc.edu.br; gustavo.dalpian@ufabc.edu.br}

\begin{abstract}
The development of spintronic devices demands the existence of materials with some kind of spin splitting (SS). In this Data Descriptor, we build a database of {\it ab initio} calculated SS in 2D materials. More than that, we propose a workflow for materials design integrating an inverse design approach and a Bayesian inference optimization.  We use the prediction of SS prototypes for spintronic applications as an illustrative example of the proposed workflow. The prediction process starts with the establishment of the design principles (the physical mechanism behind the target properties), that are used as filters for materials screening, and followed by density functional theory (DFT) calculations. Applying this process to the C2DB database, we identify and classify 358 2D materials according to SS type at the valence and/or conduction bands. The Bayesian optimization captures trends that are used for the rationalized design of 2D materials with the ideal conditions of band gap and SS for potential spintronics applications. Our workflow can be applied to any other material property.
\end{abstract}
\begin{document}

 %%%%
%%%%%%%%%%%%%%%%%%%

\flushbottom
\maketitle

\thispagestyle{empty}

\section*{Background \& Summary}

The design of materials properties usually involves two main steps: {\it i}) prediction and {\it ii}) optimization. 
The first step is typically carried out by direct calculation or experimental measurement for all possible combinations of atomic identities, composition, and structures (ACS).\cite{Zunger2018} Although this process has been successfully implemented for functionalities such as ferroelectricity~\cite{Smidt2020,PhysRevB.97.024115,Acharya2020} and two-dimensional materials~\cite{Mounet2018,Zhou2019,Gjerding_2021}, this \textit{direct approach} is usually tedious and expensive. 
The second step consists in the extrapolations of numerical correlations found with approaches such as machine learning~\cite{Schleder2019,Rodrigues2021,Schleder2019B,acosta2018analysis,Schleder2019d} or cluster expansion methods~\cite{PhysRevB.46.12587}. However, numerical relations are not necessarily transferable (i.e., limited to the set of compounds used to train the models - training set), preventing the rational design of material candidates with the optimized property out of the training set. In this work, we propose an inverse design process that integrates the rationalized prediction and optimization of materials properties as we illustrate it for the specific case of the functionality of spin splitting (SS) in 2D compounds.

Of special interest for spintronics device applications are the SS and spin polarization (SP). These functionalities are the cornerstone of spintronics, a promising, rapidly growing area that is based on the electron spin manipulation~\cite{RevModPhys.76.323, Yang2021, Acosta2018B}.  %rather than the electron charge like electronics.
%In spintronic device[], bands with SS as larger as possible and strong SP are desirable[]. 
In spintronic device prototypes, the interaction between the central region (a compound with SS bands) and electrodes (spin current source and detectors) can affect the SP properties~\cite{Acosta2018B}. Indeed, although the interface states can intrinsically possess both SP and SS, not all interfaces allow the desirable controllability of the SS. Alternative device configurations minimizing the interface effects can consist of a Van der Waals heterojunctions with a two-dimensional (2D) material as central region~\cite{Sierra2021, Xia2017, Kamalakar2017} or 2D compounds with intrinsic topological properties protecting spin currents against backscattering~\cite{PhysRevLett.122.036401, PhysRevB.94.041302,Pan2015}.
% \EO{essa última sentença achei um pouco difícil de ler.} \CM{resolvido - por favor, verificar} %However, the reported SS in 2D compounds are usually about a few meV, e.g., two order of magnitude smaller than the largest observed in 3D bulk materials[]. This could be seen as a direct consequence of little exploration of the SS in these systems. 
Unlike 3D compounds, there is no database of 2D materials with SS~\cite{MeraAcosta2020}. Thus, to illustrate the proposed design and optimization, we focus on 2D compounds possessing SS prototypes. These prototypes are historically classified by some kind of symmetry breaking (Figure~\ref{fig:SS_types}), namely: 
i) Rashba SS, with the electric breaking of inversion symmetry inducing SS~\cite{Rashba1984, Manchon2015},
ii) the Dresselhaus SS resulting from the non-electrical breaking of inversion symmetry~\cite{PhysRev.100.580}, and 
iii) the Zeeman SS induced by the breaking of the time-reversal symmetry~\cite{MeraAcosta2019, Yuan2013}.
Here, we focus only on intrinsically induced SS, i.e., the effects that are originated  from the material's intrinsic electric dipoles and spin-orbit coupling interactions, and consider a fourth additional SS prototype, namely \emph{High-Order}, as a category of observed SS effects that do not exactly follow the phenomenological SS models (see Methods).
These SS prototypes are also characterized by different spin textures, i.e., the SP pattern in the reciprocal space (Figure \ref{fig:SS_types}). \cite{MeraAcosta2021}

% \begin{figure}[ht]
% \centering
% \includegraphics[width=14cm]{new_ss_prototypes.png}
% \caption{Schematic illustration of the SS prototypes and corresponding SP patterns, where the \emph{high-order} category usually corresponds to a non-trivial spin texture.}
% \label{fig:SS_types}
% \end{figure}

We propose and implement a two-step automatic design of material properties: i) target property prediction based on the inverse design approach and ii) optimization of the target property based on Bayesian inference. 
Unlike the direct approach and machine learning, the inverse design approach requires first to define the design principles (DPs) (a physical mechanism) enabling the existence of the target properties, i.e., each of the SS prototypes above defined. 
We then develop a workflow that integrates a set of programs and scripts to automatically apply the design principles as filters, perform density functional band structure calculations including the spin-orbit coupling, and analyze the physical properties characterizing spin-polarized bands (e.g., energy position with respect to the Fermi energy, SP values, and SS magnitude) for 2D materials. We integrate this workflow to the optimization process based on Bayesian inference, in which each desirable optimization condition is maximized. Applying this approach to 2D compounds, we calculate 436 2D materials and identify more than 1200 SS at the valence and/or conduction bands, that were then classified as Rashba, Dresselhaus, Zeeman, and High-Order SS prototypes~\cite{PhysRevB.85.075404, Cartoix2006HigherorderCT}. The Bayesian analysis allowed us to find chemical and structural trends of three optimized properties relevant to spintronics: the Rashba parameter, the Zeeman spin splitting, and the band gap.

%In Figure 1, we schematically illustrate this inverse design optimization. 

%\CM{For 2D materials, stability is one of the bottle necks. ((700 words maximum)) Remember!}  

%\textbf{Description of the section (only for authors):} An overview of the study design, the assay(s) performed, and the created data, including any background information needed to put this study in the context of previous work and the literature. The section should also briefly outline the broader goals that motivated the creation of this dataset and the potential reuse value. We also encourage authors to include a figure that provides a schematic overview of the study and assay(s) design. The Background \& Summary should not include subheadings. This section and the other main body sections of the manuscript should include citations to the literature as needed. 

\section*{Methods}

We propose a workflow based on the \textit{inverse design} approach that automatically filters, selects, and designs compounds with specific functionalities\cite{PhysRevB.102.144106}, i.e., a controllable material property with potential device applications. Remarkably, in addition to the simple material prediction, commonly performed for diverse functionalities in quantum materials~\cite{PhysRevB.102.144106, Zunger2021, Malyi2020}, we integrate the proposed workflow to the automatic optimization of the target functionality based on Bayesian statistics.  

Unlike numerical correlation methods, such as machine learning and cluster expansion, the inverse design approach aims first to establish the physical mechanism (design principles) enabling the target property or functionality. The second step is to seek compounds using the design principles as filters or conditions for a rationalized design. Finally, the third step is the theoretical or experimental characterization of the magnitude of the target property for the predicted/selected compounds. In principle, unlike numerical predictive methods, in the inverse design method there is no selection of false positives, i.e., compounds that are selected but do not have the target property. The main idea of the inverse design approach is to perform an optimized materials prediction process that, compared to the "direct approach", reduces the number of necessary experiments or calculations for the target property verification. The direct approach involves the direct computation or measure of the property for all possible combinations of atomic identities, composition, and structures (ACS) attributes. 

In this section, we  describe the above-noted steps for the inverse design process in the context of spin splitting (SS) prototypes in two-dimensional (2D) materials, namely: {\bf A.} Definition of design principles, {\bf B.} Implementing design principles as filters, and {\bf C.} Target property magnitude characterization, including the SS Identification Algorithm (SSIA). Additionally, we also describe the automatic Bayesian optimization of the target properties as well as its integration with the inverse design process.

\subsection*{Automatic Inverse design approach}
\subsubsection*{A. Design principles for spin splitting prototypes}

The inverse design process starts with the definition of the physical mechanism enabling the target property. 
The application of the inverse design process for more than one target property, e.g., different spin splitting prototypes, requires to divide them into: {\bf 1.} design principles (DPs) enabling all spin splitting types, i.e., common DPs, and {\bf 2.} unique enabling DPs for each spin splitting type, as indicated in Figure~\ref{fig:DPs}. 
This design principles division is convenient for computational implementations and was also applied for the inverse design of co-functionalities by Acosta \textit{et al.}~\cite{PhysRevB.102.144106}. It is important to note that the understanding of the mechanism behind the existence or absence of SS has evolved over time, from an orthodox description based on the crystal point group symmetry to a description based on the atomic site symmetry~\cite{Zhang2014}. The latter not only reproduces the well-known description of the SS existence in compounds with non-centrosymmetric crystal point groups (e.g., Rashba and Dresselhaus SS), but also shows that compounds with local polar atomic site symmetry can have splitted split bands formed by orbital spatially localized at different material sectors~\cite{Zhang2014}. If local dipoles at different sectors cancel each other (i.e., centrosymmetric compounds), the band structure is spin degenerated. Despite the remarkable predictions and potential application of the hidden SS~\cite{Zhang2014,Liu2013,Zhang2021}, in this work, we focus on materials that explicitly have energetically discriminating spin bands in its electronic structure. In this section, we then describe the common and unique enabling design principles for the SS prototypes classification here employed.

% \begin{figure}[ht]
% \centering
% \includegraphics[width=12cm]{Desing_principles_SS.png}
% \caption{Schematic illustration of the enabling design principles for SS prototypes. 
% %\EO{Very neat design! But, as we discussed, I guess a table would be more simple to understand.}
% }
% \label{fig:DPs}
% \end{figure}

\begin{enumerate}[itemsep=0mm]
    \item \textbf{Common enabling DPs for all SS types}: The common design principles for the SS prototypes are represented in the central region of Figure~\ref{fig:DPs} (i.e., the intersection region). These include: a) non-centrosymmetry crystal symmetry, which is necessary for the SOC-induced breaking of spin degeneracy, b) non-vanishing SOC and c) non-magnetic materials, which narrow the possible materials' degrees of freedom in the analysis. As it is very well established, the breaking of both the inversion symmetry and time-reversal symmetry lifts the spin degeneracy. Our work focuses on the breaking of the inversion symmetry (i.e., non-centrosymmetric compounds - DP a).The design principles a-c are used as filters to screen materials from the original data source. Additionally, we also restricted our initial data set to compounds with a finite electronic band gap (i.e., larger than 1 meV). This is of special interest for applications in spin transistor devices.

    \item \textbf{Unique enabling DPs for SS types}:
    \begin{itemize}[leftmargin=\parindent,itemsep=0mm]
    \item \textbf{DPs for Rashba spin splitting}: E. Rashba determined that the breaking of the inversion symmetry induced by external electric fields in thin films leads to a shift in the momentum space of bands with opposite spin~\cite{Rashba1984}. This linear-in-$k$ shift near k-points preserving the time-reversal symmetry is also characterized by a splitting in the energy of the spin bands. Recently, this Rashba SS was confirmed to also exist in compounds with inversion symmetry breaking induced by the intrinsic electric dipole in 3D non-centrosymmetric bulk compounds~\cite{Maa2016, Ishizaka2011, Feng2019}. This discovery motivated the inverse design of the Rashba SS in 3D bulk compounds~\cite{MeraAcosta2020}. Although intrinsic electric dipoles can also lead to the Rashba SS in 2D materials, even without external electric fields~\cite{Zhang2014}, there is not a list of 2D Rashba compounds. To organize the design and materials' screening of this type of compounds, we first defined the enabling design principles. In addition to the common DPs previously defined, the Rashba SS in 2D compounds also requires a non-zero electric dipole (allowed by at least one polar atomic site and local dipoles that add up to non-zero, as described by~\citenum{Zhang2014}). The SS type depends on the wave vector point group~\cite{MeraAcosta2021} and hence, the DPs also include the $k$-point preserving the time-reversal symmetry, as well as a linear-in-$k$ shift. The unique DPs for the Rashba SS (i.e., non-zero electric dipoles, time-reversal symmetry, and linear-in-$k$ SS) are illustrated in the purple quadrant of Figure~\ref{fig:DPs}. We note that the \textit{Rashba SS} prototype classification employed in this work is equivalent to the R-1 classification proposed by Zhang et al. \cite{Zhang2014}, where the lack of global bulk inversion symmetry leads to non-degenerate, i.e., explicit spin split, electronic bands.
    
    \item \textbf{DPs for Dresselhaus spin splitting}:  Based on the $k\cdot p$ model, Gene Dresselhaus found that the SOC Hamiltonian describing states near a time-reversal symmetry $k$-point in a non-polar non-centrosymmetric material (i.e., with zero electric dipole) is given by $H(k)=\beta [(k^{2}_{y}-k^{2}_{z})k_{x}J_{x}+(k^{2}_{z}-k^{2}_{x})k_{y}J_{y}+(k^{2}_{x}-k^{2}_{y})k_{z}J_{z} ]$, where $J_{i}$ are the components of the total angular momentum operator~\cite{PhysRev.100.580}. In 2D compounds, the Hamiltonian is $H_{2D}(k)=\beta_{2D} (\sigma_{x}k_{x}+\sigma_{y}k_{y})$, where $\sigma_{i}$ are the Pauli matrices. This SOC term leads to a linear-in-$k$ SS. A list of 3D compounds having Dresselhaus SS has been recently reported ~\cite{MeraAcosta2020}. Based on the enabling design principles, we extend the search of Dresselhaus SS to 2D materials. Besides the common DPs, as shown in Figure~\ref{fig:DPs}, enabling DPs also include zero electric dipoles (i.e., non-polar site symmetries or polar site symmetries that add up to zero, as established in Ref.~\citenum{Zhang2014}), time-reversal symmetry, and linear-in-$k$ SS. Unlike the Rashba SS allowed by non-centrosymmetric polar point groups, the Dresselhaus SS can be found in non-centrosymmetric non-polar point groups, but it is not limited to these point groups (e.g., polar site symmetry can accidentally add up to zero). Similar to the previous case, this classification is equivalent to the D-1 effect in Ref.~\citenum{Zhang2014}.
    
    \item \textbf{DPs for Zeeman-type spin splitting}: The existence of the electron spin was first elucidated by the magnetic discrimination of states~\cite{Zeeman1897}. The breaking of time-reversal symmetry via a magnetic field, shifts in energy the Bloch bands with different spins, as in ferromagnetic compounds. More than one hundred years after Zeeman's discovery, it was established that even non-magnetic compounds can have a Zemman-type SS at $k$-points breaking the time-reversal symmetry~\cite{MeraAcosta2019, Yuan2013}. The SOC acts as an effective magnetic field that locally breaks the time-reversal symmetry but globally preserves it. The Zeeman-type SS can exist in polar and non-polar compounds; however, it satisfies the linear-in-$k$ SOC Hamiltonians describing the Rashba and Dresselhaus SSs. In this Zeeman-type SS, unlike the Dresselhaus and Rashba SS (Figure~\ref{fig:SS_types}), there are no degenerated bands. Summarizing, besides the common DPs, the Zeeman-type SS also require the local breaking of the time-reversal symmetry in the reciprocal space (i.e., wavevector point group symmetry without TR-symmetry), as indicated in Figure~\ref{fig:DPs}.
    \item \textbf{DPs for high-order spin splitting}: We define the high-order prototype as composed of SS whose band dispersion does not exactly follow the phenomenological linear-in-$k$ Rashba and Dresselhaus Hamiltonians but are degenerated at time-reversal symmetry invariant $k$-points. The design principles that apply here are polar and non-polar non-centrosymmetric crystal point groups, SS near k-points preserving the time-reversal symmetry, and non-linear SS. High-order SS is also induced by the odd terms in $k^{n}$ appearing in the SOC Hamiltonian (with $n>1$). We acknowledge that this category is broadly defined and represents a way to report materials and spin splittings that do not follow the Rashba/Dresselhaus characteristic band dispersions but still present SS that can be investigated in future theoretical works.
\end{itemize}
\end{enumerate}

Note that a given compound can present more than one SS prototype at different $k$-points, as discussed by Acosta at al.~\cite{MeraAcosta2021}, and thus the SS characterization should identify all SS types in a given compound. As both the Rashba and Dresselhaus SS prototypes share the common characteristic of having linear-in-$k$ Hamiltonians, it will be useful to label them as a \emph{Linear SS} (LSS) when implementing the SS identification algorithm, as described below, while not losing the symmetry criteria that differentiate them. Zeeman and High-order SS also have acronyms in the context of the algorithm: ZSS and HOSS, respectively). Additionally, the hidden SS can also constitute some of the potentially most promising classes of materials, which be the focus of future works limited to only hidden SP in 3D and 2D materials.

\subsubsection*{B. Materials screening based on common design principles for spin-splitting types}

In this section, we illustrate how design principles are translated into rules for compound selection, as well as the algorithms and computational implementations to perform the inverse design of SS prototypes in 2D compounds. The entire process detailed in this section is represented in the diagram shown in Figure \ref{fig:workflow_diagram}.

The starting point is the Computational 2D Materials Database (2020 version) \cite{Haastrup2018}, containing a total of 3814 unique entries generated by elemental substitution based on known 2D structural prototypes. Even though the cohesion of a given material in a two-dimensional structure after relaxation is already checked throughout the workflow performed in the database, we re-scan all the entries with a modified rank determination algorithm\cite{Larson2019}, implemented with the \texttt{analysis.dimensionality} module from Pymatgen\cite{ONG2013314}. Although not being strictly mandatory, this step is intended to unify the criteria of the symmetry and dimensionality analysis if similar procedures are applied to other two-dimensional databases that implement different strategies for 2D materials discovery. At this point, 3708 materials classified as 2D by the algorithm proceed in the screening workflow. For the list of 3708 2D compounds, the enabling DPs (See section A of Methods) are  applied as screening filters based on the materials information provided by the C2DB database, distilling the materials landscape to be analyzed by first-principles calculations.  %Table \ref{table:screening} summarizes the number of entries that proceed at each step of the screening process.
Before implementing the common DPs, we use the DFT calculated GGA-PBE band gap values in the C2DB database \cite{Haastrup2018} to remove the entries with near-zero band gap (i.e., $E_g<10^{-3}$ eV). This initial filtering results on 1020 non-zero bandgap materials. We note that throughout the DFT calculations workflow (see next subsection of Methods), the compound AuTe (identified in the C2DB with the uid: \texttt{Au2Te2-aafa8f843d5b} has a bandgap of 0.04 eV, but it is identified as metallic according to the GGA-PBE calculations implemented in the workflow of this work. This compound is then excluded from the SS identification analysis. The number of entries in this screening stage is then corrected to 1019.

For the selection of non-centrosymmetric materials (common DP-a -- see intersection region in Figure~\ref{fig:DPs}), the structure space group number is determined via the \texttt{symmetry.analyzer} module from Pymatgen \cite{ONG2013314}. Here, a restrict tolerance parameter for the space group classification is employed (\texttt{symprec = 0.001}), to determine the structure's symmetry. Such tight parameter, although being too strict for general purposes and possibly enabling the selection of false positives SS materials, is intended to select the largest group of materials that can potentially display the aforementioned SS effects in its band edges. We note, retrospectively, that if the tolerance parameter employed at this stage was the default value (\texttt{symprec = 0.01}), 21 materials that possess some type of SS in their band edges (determined by the next stages of this work) would not be identified. The number of materials that proceed at this stage of the screening process is then 500.

To guarantee the global preservation of time-reversal symmetry (common DP-b -- see intersection region in Figure~\ref{fig:DPs}), magnetic compounds are eliminated from the previous list of 500 non-centrosymmetric compounds. The time-reversal symmetry breaking can induce SS that are not necessarily induced by the spin-orbit coupling (e.g., the Zeeman effect~\cite{Tao2020} and the anti-ferromagnetic-induced SS~\cite{pekar1965combined, Naka2019}). Naturally, our approach can be extended to magnetic compounds by considering a complete analysis of magnetic point group symmetry. In this case, the SOC is not necessarily a common design principle.
Here, we use the magnetic ground state reported by the C2DB database, in which the anti-ferromagnetic configuration is restricted to duplicated unit cells. The detailed study of the magnetic ground states usually requires the DFT total energy study of multiple spin configurations and larger supercells~\cite{Kabiraj2020}. Alternatively, machine learning algorithms can be used to predict the magnetic ground states, as we demonstrated in Reference \citenum{under_constr}. This filtering process based on common DPs then results in 436 non-magnetic non-centrosymmetric semiconductors. Table~\ref{table:screening} summarizes the number of entries that proceed at each step of the screening process.

Although a large atomic SOC is usually desired, an extra filter for compounds with high atomic numbers has not been applied, so a larger set of atomic species can be analyzed in the optimization step. 
At this point, in principle, all the selected compounds can potentially have at least one SS type. Thus, the final materials selection according to the unique design principles is accompanied by the characterization of the magnitude of the spin splitting. We then proceed with the selected 436 compounds for the high-throughput calculations required to evaluate the unique enabling design principles (Figure~\ref{fig:DPs}). 

\subsubsection*{C. Target property verification: High-throughput calculations}

To classify the 436 2D compounds potentially having SS according to their SS prototype, the implementation of the unique design principles is required. This requires the calculation of all band structures. Thus, in the specific case in which the target property is the SS type, it is convenient to design an algorithm that not only filters compounds using the unique design principles according to the band shape, but also extracts the magnitude of the specific SS type. 
A workflow of \textit{ab-initio} calculations is then developed to generate band structures with a spin-polarization resolution that are analyzed in the SS identification algorithm. DFT+SOC calculations are performed using the Vienna Ab initio Simulation Package (VASP) with the projector-augmented wave (PAW) method  \cite{vasp1, vasp2} and GGA-PBE \cite{PBE} parameterization for the exchange-correlation functional. The workflow is prepared and managed using the Atomic Simulation Environment (ASE) package \cite{ase-first-paper, ase-paper}, which is integrated with VASP. %K-paths in the reduced Brillouin zone for the band structures are generated automatically through ASE, based on the unit cell structure, as proposed by Ref [Curtarolo]. % Relevant information according to parameters of the calculations is presented in section X on Suppementary Information.

%The single-particle wavefunctions are decomposed into site, orbital character, and spin projection direction in order to be analysed by the algorithm in the next step.
 
For each entry (i.e., each of the 436 selected 2D compounds), a set of three calculations is performed to i) determine the ground state charge density in a self-consistent scheme, ii) optimize the charge density in a non-collinear scf calculation, and iii) perform a band structure, non-collinear calculation (Figure~\ref{fig:workflow_diagram}). No additional relaxation procedure is performed since the available structures in the C2DB already correspond to an energy minimum. The same exchange-correlation functional is employed throughout the calculations in both the database and this work. We have verified this for a set of aleatory selected compounds at the start of the workflow. 
For all calculations, the cutoff energy for the plane-wave expansion was set to 520 eV. The choice of potentials employed in the workflow follows the Materials Project recommended PAW setup \cite{MaterialsProject}, implemented through ASE \cite{ase-paper}. Relevant parameters for the calculations i) and ii) are presented in Table \ref{table:DFT-parameters}. 
%We note that the only difference among them is the inclusion of spin-orbit coupling. 
For the sampling of Brillouin Zone according to the specific k-paths in calculation iii), a density parameter of 80 k-points per \AA$^{-1}$ was set for all entries.

 %\CM{In my opinion, it wuold be very important to have a figure illustrating the workflow (a "fluxogram"), as it is usually done in this kind of works. This potential figure can (should) be used to describe the systematic and automatic high-throughput calculations. I have modified the text, however, I still have the impression we need more details, so the readers can reproduce our calculation or at least follow the process we designed.}
 
%\subsubsubsubsection{} 
\textit{\textbf{SS Identification Algorithm (SSIA):}}
Based on the data generated from the band structures with orbital and spin resolution, an algorithm is designed to analyze the energy dispersion at the valence and conduction bands leading to the identification of the SS type (i.e.,  Rashba, Dresselhaus, Zeeman, and High-order prototypes). In this process, the unique DPs are applied as criteria used by the algorithm to evaluate and identify SSs according to i) symmetry of the $k$-point where the SS occurs, ii) band dispersion in the region of the SS, and iii) estimation of structure electric dipole. The code is built upon Pymatgen and ASE functionalities, and its underlying algorithm is detailed in this section.
The algorithm of SS identification is solely based on the band structure data, obtained by the DFT calculations. It consists of looping the analysis over the eigenvalues of the valence and conduction bands and the immediate next bands (which would represent the spin degenerated copy in a non-polarized scheme) on the high-symmetry k-points, that are labeled as time-reversal invariant momentum (TRIM) or non-TRIM k-points. For the 5 different Brillouin zones in 2 dimensions, the TRIM k-points can be directly determined and passed as a list to the SS algorithm (see section \emph{K-paths along high-symmetry lines} in the Supplementary Information - Band Structures material). Any high-symmetry k-point which is not in this list is treated as non-TRIM by the code, in the sense that it checks the possibility of having non-degenerate bands at this k-point (Zeeman SS). When analyzing the proximity of those points in each direction, three possibilities may arise:

\begin{enumerate}
    \item The pair of bands are energy degenerated on the high-symmetry k-point and in its vicinity: No SS is present;
    \item The pair of bands are not energy degenerated in a non-TRIM k-point: There is a SS gap, which is measured, and the SS is classified as belonging to the Zeeman prototype;
    \item The pair of bands is energy degenerated in the high-symmetry k-point, but not in its vicinity: A SS happens on the k-path segment between high-symmetry k-points. The SS prototype classification will then follow the characteristics of the dispersion of the bands: if both SS bands follow the same direction (the sign of the first derivative is the same in the region next to the high-symmetry k-point) the SS is immediately classified in the high-order prototype, as it do not completely obey the phenomenological Rashba/Dresselhaus linear-in-k Hamiltonian. If the bands follow opposite directions, a Rashba/Dresselhaus-type SS is observed, and the classification of the SS in one of those groups will follow the result from the structure based estimation of electric dipole. Crystal structures displaying zero (non-zero) net electric dipole are classified in the Dresselhaus (Rashba) prototype. For the last two cases, the Dresselhaus/Rashba coefficient $\alpha_{R,D}$ is computed according to Equation \ref{eq:rashba_coefficient}:
%\EO{O item 3 está um pouco confuso de entender.}
    \begin{equation}
    \alpha_{R,D} = 2\frac{\Delta E_{R, D}}{k_{R,D}}.
    \label{eq:rashba_coefficient}
    \end{equation}
    where $\Delta E_{R, D}$, $k_{R,D}$ stand for the energy difference of the spin splitted bands (SS magnitude) and the k-path interval from the SS to its correspondent high-symmetry k-point in reciprocal space, respectively. 
\end{enumerate}

Figure \ref{fig:workflow_diagram} summarizes the SS classification heuristics adopted in the algorithm. For all SS prototypes, the energy difference between the SS bands and between the SS and the VBM/CBM is computed. The result is an extensive list of SS identified at the valence and/or conduction bands for 358 materials, presented in the tables in the Supplementary Information, whose distribution according to the SS prototypes is also represented in Figure \ref{fig:distribution}. We note that single materials can have multiple, non-excluding SS prototypes at the valence/conduction bands, which are reported in the tables accordingly. 

% \begin{figure}[ht]
% \centering
% \includegraphics[width=13cm]{figures/distributions_s.pdf}
% \caption{a) Distribution of the 436 materials accordingly to different SS prototypes presented by their valence and/or conduction bands. The gray set represents the materials which our script could not identify any spin splitting; b) Swarm plot of band gap values for the entire set of materials; c) Swarm plot of Zeeman spin splitting values for ZSS materials. Inset shows the proportion of ZSS materials (239/436); d) Swarm plot of spin splittings around TRIM for the RSS/DSS and RSSH materials. Inset shows the proportion this set of materials (216/436); e) Swarm plot of Rashba parameter presented by RSS/DSS materials. Inset shows the proportion of materials presenting RSS/DSS (124/436); As a material can have multiple instances of SS in their valence or conduction bands, only the SS with the highest value are plotted on the swarm plots.}
% \label{fig:distribution}
% \end{figure}

%\EO{Gabriel, if you could put our discussion on slack about the 78 "Non-detectable SS" here}

%\begin{figure}[ht]
%\centering
%\includegraphics[width=\linewidth]{ss_bands.pdf}
%\caption{Schematic illustration of the heuristics employed to classify the identified SS according to the SS prototypes, and measurement of the Rashba/Dresselhaus coefficient ($\alpha_{R,D}$) for the respective cases. \GN{still needs some improvement}}
%\label{fig:ss_bands}
%\end{figure}

\subsection*{Inverse optimization process}

Besides the enabling (common and unique) design principles, there are other conditions that are not required for the existence of the target property, but are important for its optimization towards specific device applications. Unlike the enabling design principles, the optimizing design principles are not necessarily physical mechanisms, but characteristics related to the chemical compositions and structure (that do not affect the existence of the target property). For SS and SP prototypes, for instance, one would like to have a compound with i) a large enough band gap to allow gate-controllable position of the Fermi energy, and ii)  effective masses set to increase the charge carrier mobility to provide control over the performance of semiconductor devices. Additionally, it is also desirable to optimize some other properties characterizing SS prototypes and the efficiency of spintronic devices, namely: iii) large SS (i.e., larger than 25 meV), iv) large SP coefficient (i.e., larger than 1.3 meV), and v) position of the SS with respect to the Fermi energy.    

In an ideal scenario, properties i-v should have optimized values. However, these optimum values could be physically contradictory according to the usual trends defined by the chemical composition. For instance, while large SS are usually expected in compounds formed by atoms with large atomic numbers, large bandgaps tend to be found in compounds formed by atoms with small atomic numbers. This suggests that large bandgaps and large SSs are in some sense contradictory. The question arising  from these apparent contradictions is: How to find the optimal candidate? This problem is also evident in other areas. For instance, the apparent contradiction between a high thermal insulation and electrical conductivity, which is desirable for thermoelectric applications. Here, we illustrate a strategy to address this problem for the specific case of compounds with SS prototypes for spintronic device applications, as shown in section Technical Validation. As we explain below, the inverse optimization process is based on Bayesian statistics using the materials' composition and crystal structure.

%atomic properties (e.g., atomic number, atomic composition,...\CM{complete}) as independent variables.

\subsubsection*{Bayesian statistics for materials property optimization} 

Bayesian Inference (BI) is a statistical method of inference based on Bayes' Theorem. The Bayes' Theorem specifies how one should update the probabilities when new  information is given. Starting from a hypothesis $H$, such as a material belonging to class $h$, and a property $A$ of this material, one can define $P(H|A)$, i.e., the probability of an material belonging to h given that it has the property A. This is called posterior probability and it is given by the Bayes Theorem:

\begin{equation}
P(H|A) = \frac{P(A|H)}{P(A)}P(H)
  \label{eqn:bayes}
\end{equation}

\noindent where the three right-side terms are:
\begin{itemize}
    \item $P(A)$: Probability of a material having the $A$ property;
    \item $P(H)$: The prior probability. This is the probability of $H$ without knowing anything regarding the material. It is simply given by the ratio of materials belonging to $h$;
    \item $P(A|H)$: The likelihood function. The probability of a material with property $A$ given that it belongs to $h$.
\end{itemize}

 Equation~\ref{eqn:bayes} shows how to update the probability of $H$ given new information regarding $A$. Observing how the knowledge of a material's property $A$ updates the prior probability $P(H)$ allows to infer about correlations between $A$ and $h$. Therefore, an increasing probability update shows a strong correlation between $A$ and $h$.

The likelihood term $P(A|H)$ is a calculated probability given by feature/property $A$ and its distribution in materials within $h$. Properties can be continuous, discrete, or categorical, which implies that different probability distributions should be used accordingly to $A$'s nature. 

For the case of $A$ as a categorical feature, such as for structural cluster, we are interested in the posterior probabilities given that the material belongs to a structural cluster $t$ from the $n$ possible structural clusters, i.e. we want to calculate $P(H|A = t)$. One should use the Categorical distribution for the likelihood term $P(A = t|H)$, given by:

\begin{equation}
P(A = t|H) = \frac{N_{th} + \alpha}{N_h + n\alpha},
  \label{eqn:categorical}
\end{equation}

\noindent where $N_{th}$ is the number of times the materials from structural cluster $t$ appears in the class $h$, $N_h$ is the total number of materials belonging to the class h, and $\alpha$ is a regularization term. The regularization term prevents $P(A = t|H)$ from having absolute values (0 or 100\%) when the category $t$ has all materials outside or within the class $h$. 

%This is useful for regularizing structural clusters with few materials.

% For the case of $A$ as the presence of an element as a cation/anion in the material's composition, $A$ can assume False or True values. We wish to calculate $P(H|\text{X element as cation} = True)$ and $P(H|\text{Y element as anion} = True)$ and, due the boolean nature of these features, one should use the Bernoulli distribution:

For the case of $A$ as the presence of an element in the material's composition, $A$ can assume False or True values. We wish to calculate $P(H|\text{X element in composition} = True)$ due to the boolean and non-exclusive nature of this feature (a material can have more than one element in its chemical composition). One then should use the Bernoulli distribution:

\begin{equation}
P(A_X|H)=\left\{\begin{array}{ll}
p_X & \text { if } A_X=True \\
1-p_X & \text { if } A_X=False
\end{array}\right.,
\label{eqn:bernoulli}
\end{equation}

% where $A_X$ is the event of having element X as cation/anion and $p_X$ is similar to the probability given by Equation~\eqref{eqn:categorical}:

where $A_X$ is the event of having element X in its composition and $p_X$ is similar to the probability given by Equation~\ref{eqn:categorical}:

\begin{equation}
p_X = \frac{N_{X=True,h} + \alpha}{N_h + \alpha},
  \label{eqn:p}
\end{equation}

where $N_{X=True,h}$ is the number of materials with X in the composition that belongs to class $h$, and $N_h$ the total number of materials belonging to class $h$. Once again, $\alpha$ acts as a regularization term.

The Categorical and Bernoulli distributions were used to infer the effect of crystal structure and composition on the calculated spin-splitting properties, respectively, as presented in the Technical Validation Section. We used the scikit-learn~\cite{sklearn} implementations, with $\alpha=1$ as regularization.

\section*{Data Records}

As each material may present multiple SSs in its band structure (in the valence and/or conduction band) classified into different SS types, the generated data shows a highly unstructured nature. For this reason, we opted to provide the complete data in multiple formats, suited for the different types of data structure and use cases.

Firstly, an overview of this work's main findings is contained in tables available in the Supplementary section and in an Excel-compatible .csv file. There, the reader may find the full list of SSs identified in this work, separated by the SS prototypes (Rashba, Dresselhaus, Zeeman, and High-order) (See \emph{Design principles for spin splitting prototypes} section), as well as the materials general information (e.g. id, symmetry, band gap, energy above convex hull) and SS related information (e.g. SS magnitude and localization in the band structure). As each line in the table represents a single SS, one compound may be repeated several times in each SS category. 
For a visual representation of the data, the Supplementary section also contains a pdf file with the rendered image of the structure and band structure with spin polarization resolution for all materials calculated in this work. The reader is then able to correspond the SS reported in the tables with its localization in the plotted band structure.

These images are also available in the Materials Cloud \cite{Talirz2020} repository for this work \cite{materials_cloud_data}, subdivided into folders for each compound. There, the reader also finds a .cif file containing the material’s crystal structure information, that determines the choice of the unit cell and origin of the coordinate system of the structure representations employed in the calculations.

Regarding the raw results of the calculations, these are available in two sources. For data provenance, we store the main input and output files of the VASP DFT band structure calculations in a NOMAD repository ~\cite{nomad_data}. In this manner, the reader can find the necessary raw files that would reproduce the calculations in this work. For accessing the band structure results, on the other hand, we alternatively provide a DFT-code-agnostic format which mainly relies on Pymatgen \cite{ONG2013314} and ASE \cite{ase-paper} python objects. These are available in a pickle file inside each compound’s folder in the Materials Cloud repository \cite{materials_cloud_data}. There, the reader may get the full list of k-point coordinates, eigenvalues, orbital projections, spin polarization, and other relevant data of all the band structure calculations in this work. The README.md file available in the repository contains detailed instructions to open the data using Python.

Regarding the main results of this work, which represent the full post-processed data with the SS identification and description generated by the developed algorithm for all screened materials, these are available in a single dataset as a JSON file in the Materials Cloud repository. Alternatively, we also made it available as a binary export dump file that can be imported directly to a MongoDB database (\url{www.mongodb.com}). These two files contain the same information and may suit different preferences and use cases. Detailed instructions for accessing both data files are also available in the README.md file.

In this dataset, each entry corresponds to one material and provides three classes of information: the material-specific data, the band structure description data, and the spin-splitting description data.
 
The former provides the material's general description regarding its composition, crystal structure, band gap, and other properties. Box~\ref{lst:material_keys} shows a complete overview of this data. The band structure data (Box~\ref{lst:bands_keys}) gives information about the band structure calculations, such as the number of calculated bands, the number of k-points, the Brillouin zone, and all energy eigenvalues, 
% in addition to the pickle file in each materials folder. 
Here, the key \texttt{NOMAD\_files} has the URLs to the material-specific files for the band structure calculation at the NOMAD repository. 

The spin splitting data, as the name suggests, provides information of all spin splittings occurring in the material's valence and conduction band, which compiles all the data resulting from SSIA into SS types (LSS, ZSS, or HOSS) and their specific parameters. Box~\ref{lst:ss_keys} details all keys describing the SSs. 

The user can easily query the data by importing it to MongoDB (check the MongoDB website on setting up a database in your specific system) or any other NoSQL database option. Alternatively, one can also convert the JSON file to a simple table, with special care to normalize the \texttt{vb}/\texttt{cb} keys and their subkeys. By using the MongoDB approach, one can query the unstructured spin splitting data for further analysis by using the many programming languages which MongoDB provides support, or by using the mongosh (shell environment) or even the Compass program (MongoDB's graphical user interface environment). In the README.md file from the Materials Cloud repository we also provide instructions for doing so. As an example, the simple query shown in Box~\ref{lst:query} retrieves the cations, anions, and the structural cluster of all compounds with a Zeeman spin splitting greater than 0.1 eV in the valence or the conduction band. The resulting data will be used in the next section of Technical Validation.

\includepdf[pages={1-4}]{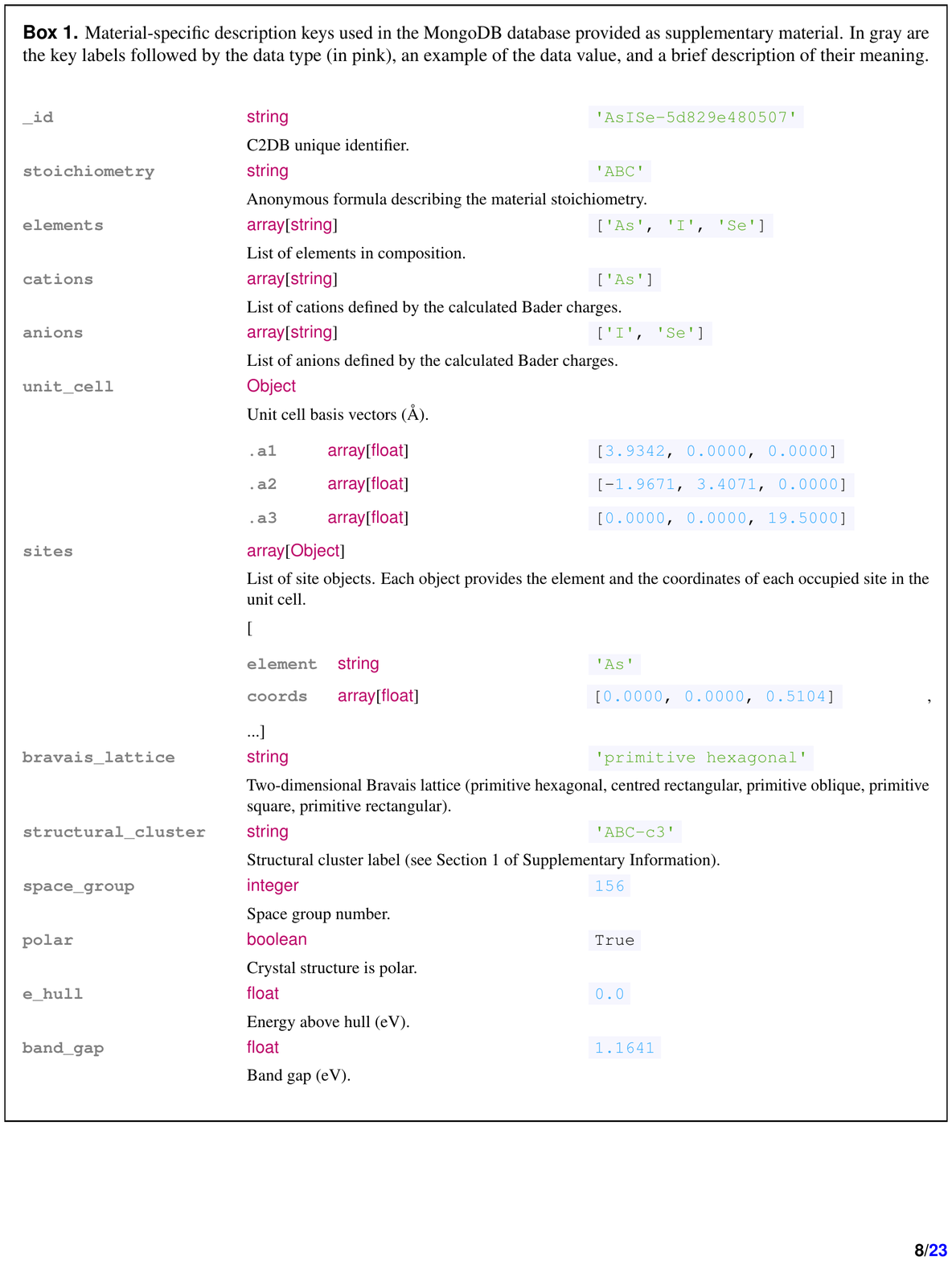}

\section*{Technical Validation}

%\GD{Gabriel has to open a sub-section in this section and discuss the anti0-crossing algorithm and insert the figures of anti-crossing}

The data generated in this work opens the way for two possible direct applications: i) the selection of optimized compound, i.e., \textit{the best of a class} for a given application and ii) the understanding of the interplay between the physical mechanisms behind a given target property based on data trends that can be analyzed with the Bayesian optimization of materials properties. We then highlight in this technical validation section how the proposed workflow for selection and optimization leads to the \textit{best of a class} for a 2D semiconductor with large SS, illustrating the use of the Bayesian optimization applied for the specific case of the Zeeman SS.

\subsection*{Bayesian inference}

% We used the generated data to analyse the effects of composition and crystal structure on SS magnitude and type. In this analysis, we employed the Bayesian inference explained previously.

% , which is a very simple statistical inference method based on Bayes' Theorem. 

As symmetry and structure are major components of bands behavior, it is fundamental that we convey these properties into data in an appropriate way. The C2DB provides a label called crystal prototype, with a \textit{stoichiometry} - \textit{space group} - \textit{occupied Wyckoff positions} format. The H-phase of monolayer \ce{MoS2} is described as \ce{AB2}-187-ai, for example. We have observed that grouping the materials using this label leads to some erroneous clusters. Some materials that should be clustered together are separated, resulting in a sparse space of crystal prototypes. Figure 1 in Supplementary Information illustrates this and some other limitations of using space group or crystal prototype for grouping materials. 
To bypass these limitations and also to get to more intuitive conclusions regarding the structure degree of freedom, we used the methodology described in reference~\citenum{zimmermann2020local} to generate crystal fingerprints (CF) for each of the materials in the C2DB. As the resulting CFs have large dimensionality, we then used them as input for the UMAP~\cite{umap} embedding technique. The result is a 2d embedding that characterizes the structural differences and similarities of C2DB's materials. We used DBSCAN~\cite{dbscan} to define the clusters in the embedding (Figure 2 in Supplementary Information) and then further subdivided them accordingly to the materials' stoichiometry. As an example, \ce{MoS2} is labeled as \ce{AB2}-c22 because of its binary stoichiometry and its given cluster number (22). A total of 23 structural clusters were found for the 436 2D materials. Further details of this process are presented in the Section~1 of Supplementary Information.
Due to the categorical nature of the structural cluster (i.e., the material can only be at one specific cluster), we used the categorical distribution implementation of Naive Bayes for evaluating how being a member of such clusters weights for a given material regarding a target property.

As an illustrative example, we used the Zeeman spin-splitting (ZSS) as a target property to be analyzed. Setting a bottom limit of 100 meV for materials with "large ZSS", we then proceeded to use Bayesian inference to generate probabilities for each structural cluster. These probabilities are the posterior probabilities, i.e., the probability that a material has a large ZSS given that it is in a given structural cluster. They were calculated using the categorical distribution given by Equation~\ref{eqn:categorical}. The results are presented in Figure \ref{fig:structure}. Only structural clusters with at least 5 materials are shown and analyzed.

% \begin{figure}[ht]
% \centering
% \includegraphics[width=16cm]{figures/structure_figure_s.pdf}
% \caption{Top: Bayesian probability of having a Zeeman spin splitting greater than 100 meV given a structural cluster. Only the structural clusters with at least 5 materials are shown. The prior probability is the ratio of the 436 materials with the target property (28\%); Bottom: The structural clusters with highest posterior probabilities.}
% \label{fig:structure}
% \end{figure}

The posterior probabilities shows that the only structural clusters associated with the target are \ce{AB}-c25 (Buckled Hexagonal AB), \ce{ABC}-c3 (T-TMDC MXY Janus), \ce{ABC}-c4 (H-TMDC MXY Janus), and \ce{AB2}-c4 (H-TMDC). All of them are hexagonal and present ZSS at the $K$ points of the Brillouin Zone.
The Zeeman-type spin splitting in these structures, except for ABC-c3, has been extensively studied in the context of valleytronics~\cite{schaibley2016valleytronics}. In \ce{AB2}-C4 (H-TMDCs), the non-degeneracy of spin states at $\pm K$-points combined with time-reversal symmetry requires that spin splittings at $+K$ and $-K$ to be opposite~\cite{xiao2012coupled, jones2013optical}. This condition gives rise to the called valley Zeeman effect~\cite{aivazian2015magnetic, srivastava2015valley, macneill2015breaking, li2014valley}, which happens when the degeneracy between valleys $+K$ and $-K$ is broken by applying a perpendicular magnetic field. The same characteristic can be found in the literature for ABC-c4 (H-TMDC MXY Janus) systems~\cite{peng2018valley, hu2018intrinsic, cheng2013spin} and AB-c25 (Buckled Honeycomb AB) structures~\cite{di2015emergence}, with the latter presenting ferroelectricity as a co-functionality. 
For ABC-c3 (T-TMDC MXY Janus), the spin-splitting arises uniquely from the intrinsic out-of-plane dipole originated from the electronegativity difference between X and Y anions. Materials with this structure have been studied primarily because of their large Rashba-type spin splittings located around $\Gamma$~\cite{ma2014emergence, zhuang2015rashba, riis2019classifying}. The ZSS localized at $K$ shows to be inaccessible as it is usually distant from the VBM/CBM around $\Gamma$.

To evaluate the composition influence on spin splitting properties, we first separated the composition of each material into cations and anions. This separation is performed according to the calculated Bader~charges~\cite{bader} of each ion in the unit cell, which is currently available in C2DB. If the cation/anion X is (not) present at the material composition, the feature cation/anion X is set to (False) True. The distribution of cations and anions in the composition of all 436 materials is shown in Figure 5 from Supplementary Information. We used a Naive Bayes implemented with the Bernoulli distribution (Equation~\ref{eqn:bernoulli}) as these features are not exclusive, i.e., one entry can have more than one kind of cation/anion. The results are presented in Figure \ref{fig:composition}. Only the cations/anions present in at least five materials are shown.
We can see that the cations have a larger range of posterior probabilities, indicating that they are more important than anions regarding the Zeeman spin splitting in those systems. One can also notice the top-down trend in the periodic table groups: heavier atoms are associated with higher ZSS due to their higher spin-orbit coupling. The most significant cations were \ce{Bi}, \ce{Sb}, \ce{As}, \ce{Ir} and \ce{Hf}, while the most important anions were the heaviest ones: \ce{Te} and \ce{I}.

% \begin{figure}[ht]
% \centering
% \includegraphics[width=\linewidth]{figures/cations_anions_figure_s.pdf}
% \caption{Bayesian inference heatmap for Zeeman spin splitting greater than 100 meV given the presence of an element as a cation/anion in material's composition. Each element cell also shows its posterior probability and its count (i.e. the number of materials with the element as cation/anion). The prior probability is 28\%.}
% \label{fig:composition}
% \end{figure}

This same methodology is then applied to Rashba/Dresselhaus spin-splitting, Rashba parameter, and band gap as target properties. The results of the Bayesian Inference analysis are provided by Table~\ref{tab:results}. The analogous heatmaps are presented in the Supplementary Information (Section 4 - Target properties structural and compositional heatmaps).

\subsection*{Case study for optimizing Zeeman spin splitting}

Combining the optimal compositions and structures found in the previous section might lead to the exploration of novel spin-based materials by simple ion substitution. We used the results in Table~\ref{tab:results} for ZSS and combined the optimal structural clusters, cations, and anions to generate a set of materials that are likely to show large ZSS (defined above as larger than 100 meV). The result is a set of 30 materials, from which nine were already in the screened dataset of 436 non-metal materials (Duplicated NM), seven were in the general C2DB database as metals (Duplicated M), and 14 are new combinations (New). Figure~\ref{fig:dist_zss}a illustrates the combinatorial aspect of this ZSS optimization.

The new ion-substituted combinations were then structurally relaxed until the Hellmann-Feynman forces on each ion were less than 1.0 meV/Å, and then their respective band structures were calculated. All of them maintained their structural characteristics. Also, all of them presented a ZSS larger than 100 meV, the target of the optimization process. However, except for two materials (\ce{IrTeI} on the two Janus phases), almost all of them resulted in metallic band structures. The set Duplicated M consisting of 7 metals in C2DB that fits the criteria also validates our optimization process, except for only one material with ZSS below the threshold. Figure~\ref{fig:dist_zss}b shows the ZSS distribution over these different sets of materials.

% \begin{figure}[ht]
% \centering
% \includegraphics[width=\linewidth]{figures/zss_dist_s.pdf}
% \caption{a) Illustration of combination of optimized structural clusters and optimized compositions; b) Strip plot of Zeeman spin splitting. Blue circles correspond to all Zeeman spin splittings in the 436 materials dataset (materials that do not have ZSS are represented by a Zeeman spin splitting of zero). Red circles: Duplicated M; Diamonds: New materials generated by optimizing criteria; White diamonds: New materials which are gapped.}
% \label{fig:dist_zss}
% \end{figure}

\subsection*{Best of a class: the case of Zeeman SS Materials} 

% \begin{figure}[ht]
% \centering
% \includegraphics[width=10cm]{figures/temp.pdf}
% \caption{\EO{Temporario}}
% \label{fig:pareto}
% \end{figure}

To present a list of "best of a class" materials from the resulting screened C2DB database, one first needs to specify what "best" means with well-defined objectives. For such, we delineated three parameters:

\begin{itemize}
    \item \textbf{SS intensity}: Given by the energy delta of spin splitting. It is defined as Zeeman spin splitting for non-TRIM points and as Rashba/Dresselhaus spin splitting for TRIM points;
    \item \textbf{SS accessibility}: The energy delta of the spin splitting to the VBM or CBM. It measures how accessible the SS is for exploitation and experimental verification; 
    \item \textbf{Energy above hull (ehull)}: It measures the material stability to its decomposition to other structures with the same composition. An ehull of 0 eV means that the material is thermodynamically stable in comparison to other phases.
\end{itemize}

We described the dataset in terms of these three parameters so one can obtain materials with a large and accessible SS but, at the same time, that are thermodynamically stable in comparison to other competing phases. By filtering out the materials with ehull > 30 meV, with accessibility < 100 meV and Zeeman spin splitting > 30 meV, we obtained the materials presented at Table~\ref{tab:best_zss}.

\subsection*{Effect of anti-crossing bands}

Acosta \textit{et al.}.  were the first to propose a causal relationship between the anti-crossing orbital character of energy bands and a large Rashba coefficient for 3D-bulk compounds \cite{MeraAcosta2020}. In this work, we investigate the extension of this concept in the context of two-dimensional materials. 

An anti-crossing analysis procedure is implemented in the SSIA and consists of two steps: i) identifying aligned pairs of SS between valence and conduction bands and ii) measuring the change in the orbital character of the selected bands in the k-interval between the high-symmetry k-point and the inflection point where the SS is maximum. If a monotonic and inverse change on the orbital contribution is present in the pair of SSs at the valence and conduction bands, the anti-crossing of SS bands (AC-SS) is verified. 

Figure \ref{fig:anticrossing_figure} a) illustrates this analysis for the 2D compounds with SS identified in this work. In such context, the presence of anti-crossing bands may indicate a sufficient condition, as all AC-SS present Rashba coefficients larger than 0.862 eV/\AA$^{-1}$. But they are not necessary, as far as other large Rashba compounds do not display a measurable anti-crossing between valence and conduction bands. While we understand that other symmetry-driven physical mechanisms can play an important role in SS in 2D compounds, an indirect effect of anti-crossing bands is also noted, as it is illustrated for the \ce{AsIS} (MXY Janus) compound in Figure \ref{fig:anticrossing_figure} b). For its respective band structure, anti-crossing is verified among the valence and other deeper bands, leading to a large Rashba coefficient for such SS that may have a consequential effect on the SS observed at the valence band.

\section*{Usage Notes}

As noted in the Data Records section, the complete data of SS for all materials obtained through this work are available in a .csv dataset, within columns listing the SS effects classified across the different SS prototypes. The contents within each cell are in the form of nested python dictionaries, so python coupled with data analysis libraries, e.g. \texttt{pandas}, are the recommended tools to access the data in order to filter for materials of interest.

The computational code, i.e. a python class, used for all the SS analysis from the output of the DFT calculations is also provided (see Code Availability section) and can be used to identify and measure SS effects of other 2D materials calculations. Due to the context of this work, the algorithm currently supports non-collinear band structure calculations performed with VASP, whose k-paths over the reciprocal space have been generated with ASE's Brillouin zone sampling automatic scheme. The code initialization requires only the specification of the folder path where the band structure calculation was performed, and presents different methods to identify SS effects in materials valence/conduction bands and the presence of anti-crossing bands, and also offers tools for plotting band structures with spin texture resolution. A more detailed description of the current functionalities of the code are available in its GitHub repository.

%\textcolor{gray}{The Usage Notes should contain brief instructions to assist other researchers with reuse of the data. This may include discussion of software packages that are suitable for analysing the assay data files, suggested downstream processing steps (e.g. normalization, etc.), or tips for integrating or comparing the data records with other datasets. Authors are encouraged to provide code, programs or data-processing workflows if they may help others understand or use the data. Please see our code availability policy for advice on supplying custom code alongside Data Descriptor manuscripts.}

%\textcolor{gray}{For studies involving privacy or safety controls on public access to the data, this section should describe in detail these controls, including how authors can apply to access the data, what criteria will be used to determine who may access the data, and any limitations on data use. }

\section*{Code availability}

The entire computational code employed in the SS analysis within this work is openly available at the GitHub repository \url{github.com/simcomat/SS_2D_Materials}. It is intensely built upon tools and methods from Pymatgen\cite{ONG2013314} and ASE \cite{ase-paper}, and provide functions to identify, measure and classify SS effects that appear valence/conduction bands of 2D materials band structure calculations.

\bibliography{sample.bib}

\section*{Acknowledgements} 

The authors thank Brazilian agencies FAPESP (2017/02317-2, 2018/11641-0, 2018/11856-7 and 2019/04176-2) and CNPq for financial support. High throughput calculations were performed using the Santos Dumont supercomputer at LNCC. 

\section*{Author contributions statement}
GMN performed the ab initio calculations, wrote the code that calculated the spin splittings and generated band structures and datasets. EO performed the inverse optimization process, with the Bayes analysis and organized the MongoDB database. CMA and GMD conceived the idea. AF, CMA and GMD supervised the work. All authors helped writing and reviewing the manuscript.

\section*{Competing interests}
The authors declare no competing interests. 

\section*{Figures \& Tables}

\begin{figure}[ht]
\centering
\includegraphics[width=14cm]{new_ss_types.pdf}
\caption{Schematic illustration of the SS prototypes and corresponding SP patterns, where the \emph{high-order} category usually corresponds to a non-trivial spin texture.}
\label{fig:SS_types}
\end{figure}

\begin{figure}[ht]
\centering
\includegraphics[width=12cm]{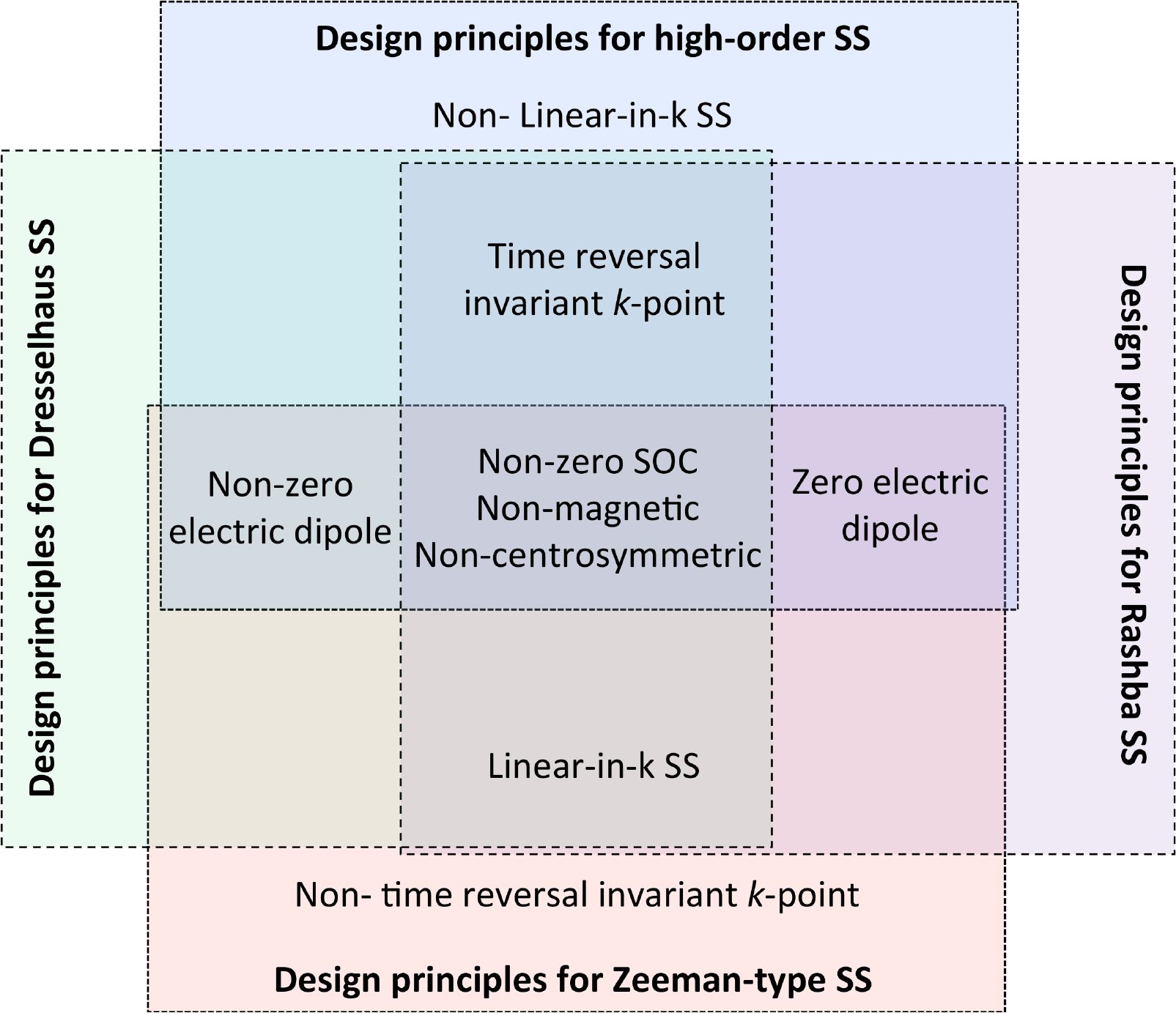}
\caption{Schematic illustration of the enabling design principles for SS prototypes.}
\label{fig:DPs}
\end{figure}

\begin{table}[h!]
\centering
\begin{tabular}{|c c|} 
 \hline
 Filter & Entries \\ [0.2ex] 
 \hline
 2D algorithm\cite{Larson2019} & 3708 \\ 
 PBE Gap > 0 & 1019 \\ 
 Non-centrosymmetric structure & 500 \\ 
 Non-magnetic & 436 \\ 
 \hline
\end{tabular}
\caption{Number of entries in each screening step based on the enabling design principles.} 
\label{table:screening}
\end{table}

\begin{table}
\centering
\begin{tabular}{|c c|}  
 \hline
 k-point mesh & $\Gamma$-centered, 11x11x1 grid \\ [1ex] 
 Energy difference for electronic convergence & 1.0 $\times$ 10$^{-6}$ eV \\ [1ex] 
 Smearing & tetrahedron method with Blochl corrections (ISMEAR = -5) \\ 
 \hline
\end{tabular}
\caption{Relevant parameters for the DFT calculations i) and ii) in the workflow performed for each entry.}
\label{table:DFT-parameters}
\end{table}

\begin{figure}[ht!]
    \centering
    \includegraphics[width=0.85\textwidth]{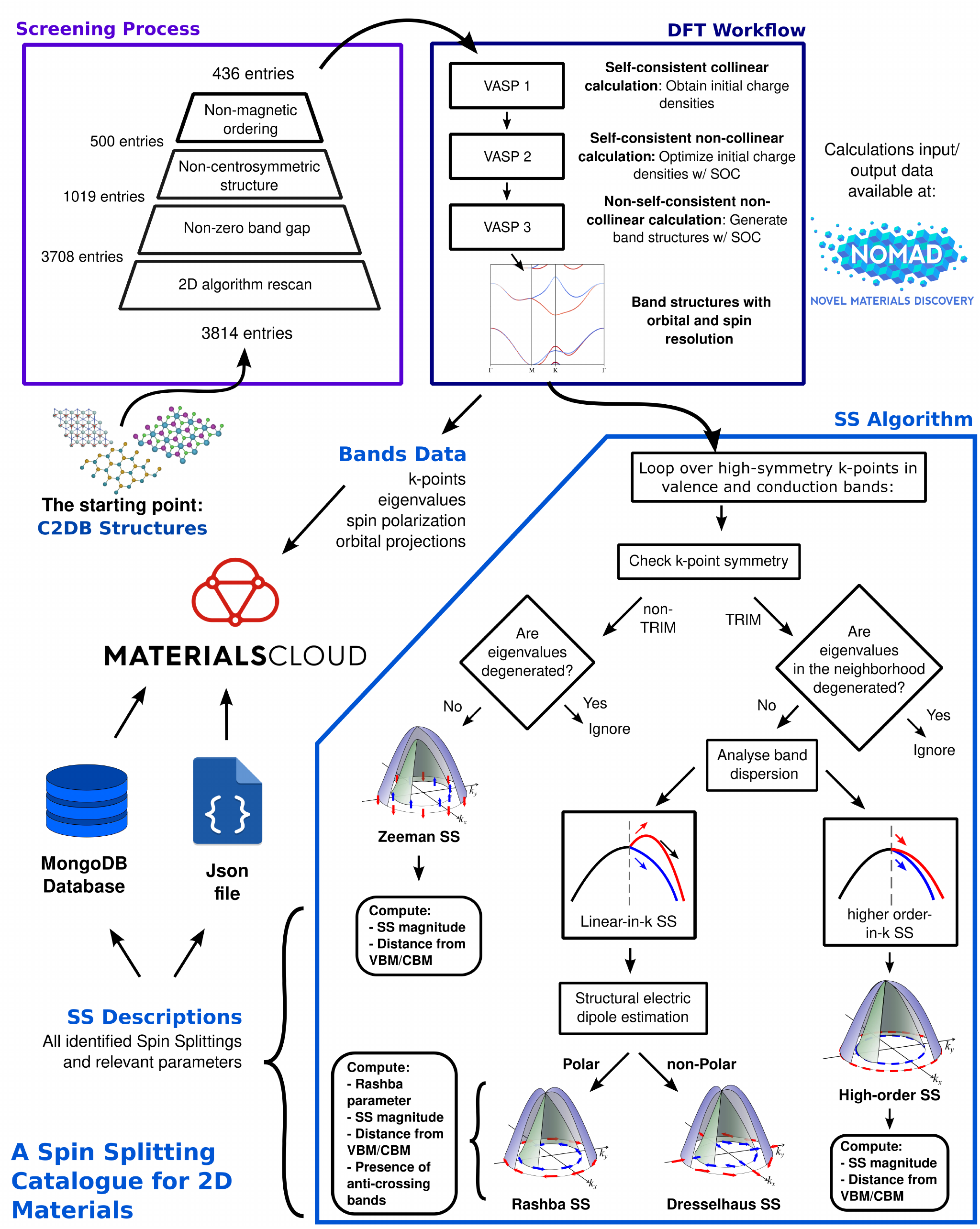}
    \caption{Representation of the workflow outlined in section B of Methods. It starts with the 3814 materials from the C2DB (2020 version) and proceeds to the screening process (purple rectangle), sequence of DFT calculations (dark blue rectangle), and SS identification (blue rectangle), where its underlying algorithm is schematically represented. The final result is a database of SS in 2D materials, identified in valence and/or conduction bands, available in multiple formats and data types (see \emph{Data Records} section).}
    \label{fig:workflow_diagram}
\end{figure}

\begin{figure}[ht]
\centering
\includegraphics[width=13cm]{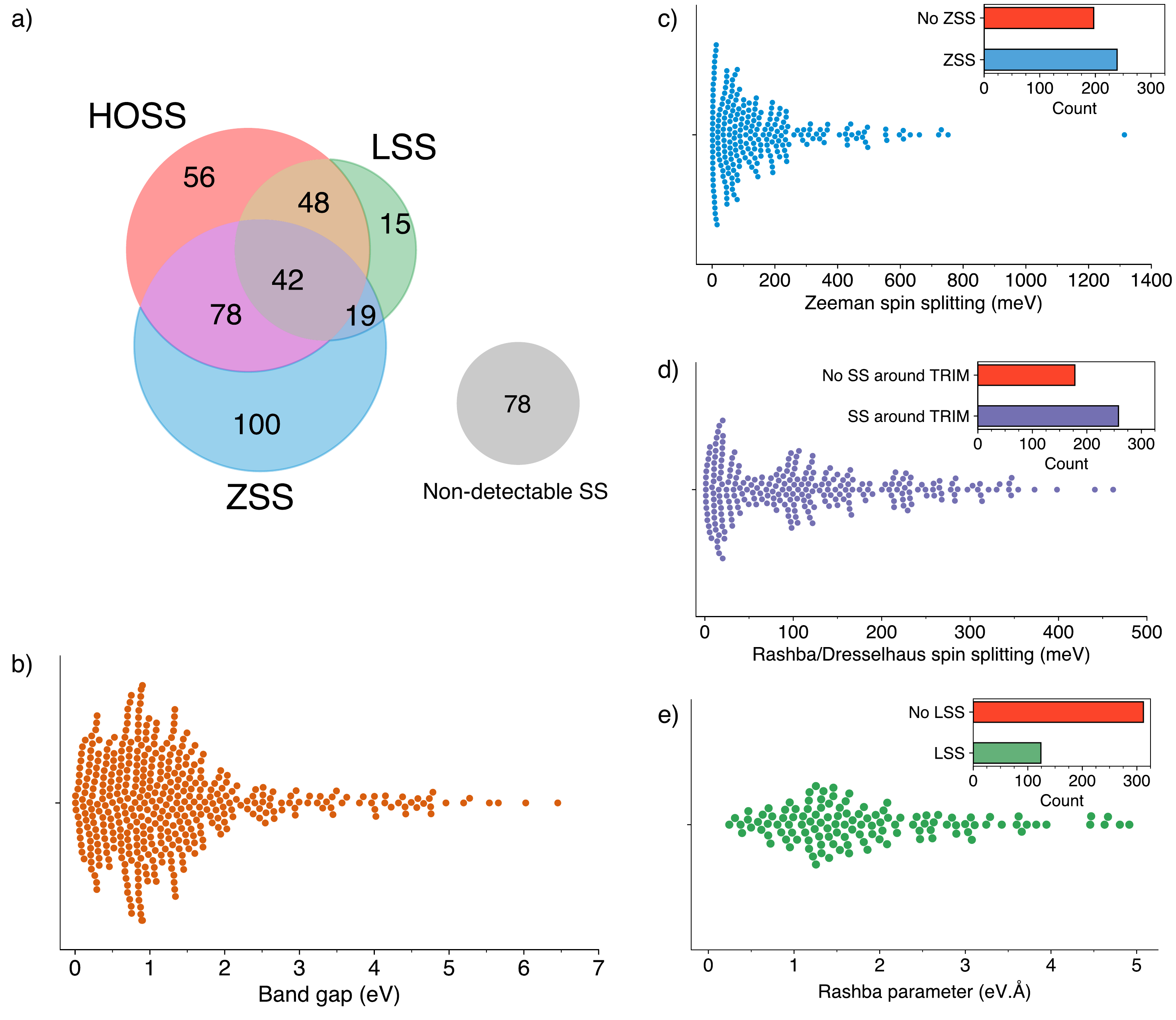}
\caption{a) Distribution of the 436 materials according to different SS prototypes presented at their valence and/or conduction bands. The gray set represents the materials in which the SS Identification Algorithm (SSIA) could not identify any spin splitting; b) Swarm plot of band gap values for the entire set of materials; c) Swarm plot of the Zeeman spin splitting (ZSS) values. Inset shows the proportion of ZSS materials (239/436); d) Swarm plot of spin splittings around TRIM k-points for the LSS and HOSS materials. Inset shows the proportion in this set of materials (216/436); e) Swarm plot of the Rashba parameter presented by LSS materials. Inset shows the proportion of materials presenting LSS (124/436); As a material can have multiple instances of SS in their valence or conduction bands, only the SS with the highest value is plotted on the swarm plots.}
\label{fig:distribution}
\end{figure}

% \begin{table}[h!]
% \begin{center}
% \begin{tabular}{|c c m{7cm}|} 
%  \hline
%  Key & Available for the SS prototypes & Description \\ [0.5ex] 
%  \hline\hline
%  \texttt{'label'} & LSS / ZSS / HOSS & High-symmetry k-point label \\ 
%  \hline
%  \texttt{'direction'} & LSS / HOSS & Direction of the k-path segment analyzed by the algorithm in the vaicinity of the high-symmetry k-point \\ 
%  \hline
%  \texttt{'rashba\_param'} & LSS & Measured Rashba coefficient \\ 
%  \hline
%  \texttt{'spin\_splitting'} & LSS / ZSS / HOSS  & Measured SS magnitude \\ 
%  \hline
%  \texttt{'accessibility'} & LSS / ZSS / HOSS  & Energy difference from the SS to the VBM/CBM of its corresponding band \\ 
%  \hline
%  \texttt{'anti\_crossing'} & LSS / HOSS  & Presence of anti-crossing bands among aligned SS across valence and conduction bands\\ 
%  \hline
% \end{tabular}
% \caption{SS parameters automatically computed with the SS algorithm developed in this work. Both the data generated for all materials and the computational code are available for public access (see Data Records and Usage Notes section).}
% \label{tab:ss_parameters}
% \end{center}
% \end{table}

\begin{figure}[ht]
\centering
\includegraphics[width=16cm]{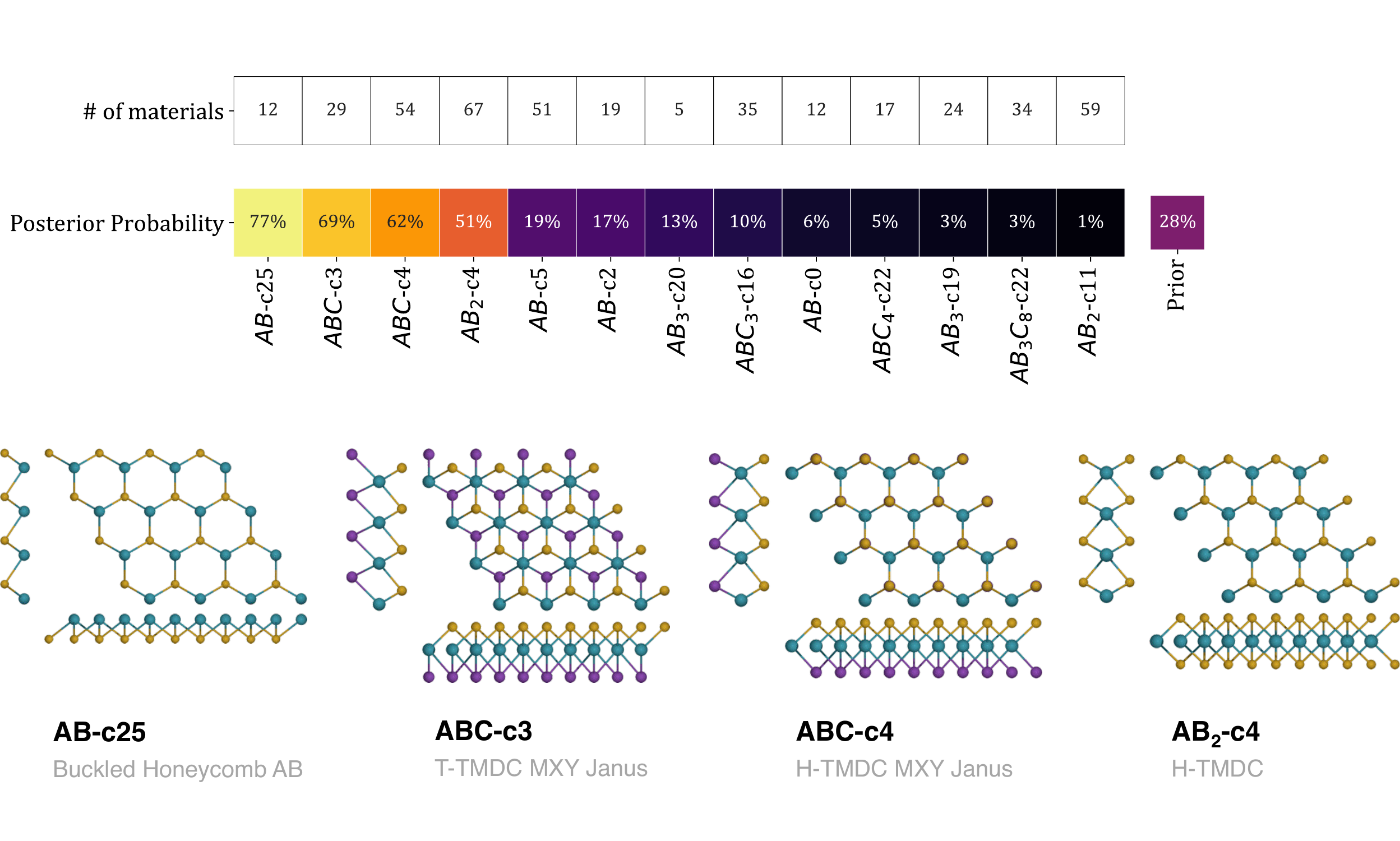}
\caption{Above: Bayesian probability of a material to have a Zeeman spin splitting greater than 100 meV given a structural cluster. Only the structural clusters with at least five materials are shown. The prior probability is the ratio of the 436 materials with the target property (28\%). Below: The structural clusters with the highest posterior probabilities.}
\label{fig:structure}
\end{figure}

\begin{figure}[ht]
\centering
\includegraphics[width=\linewidth]{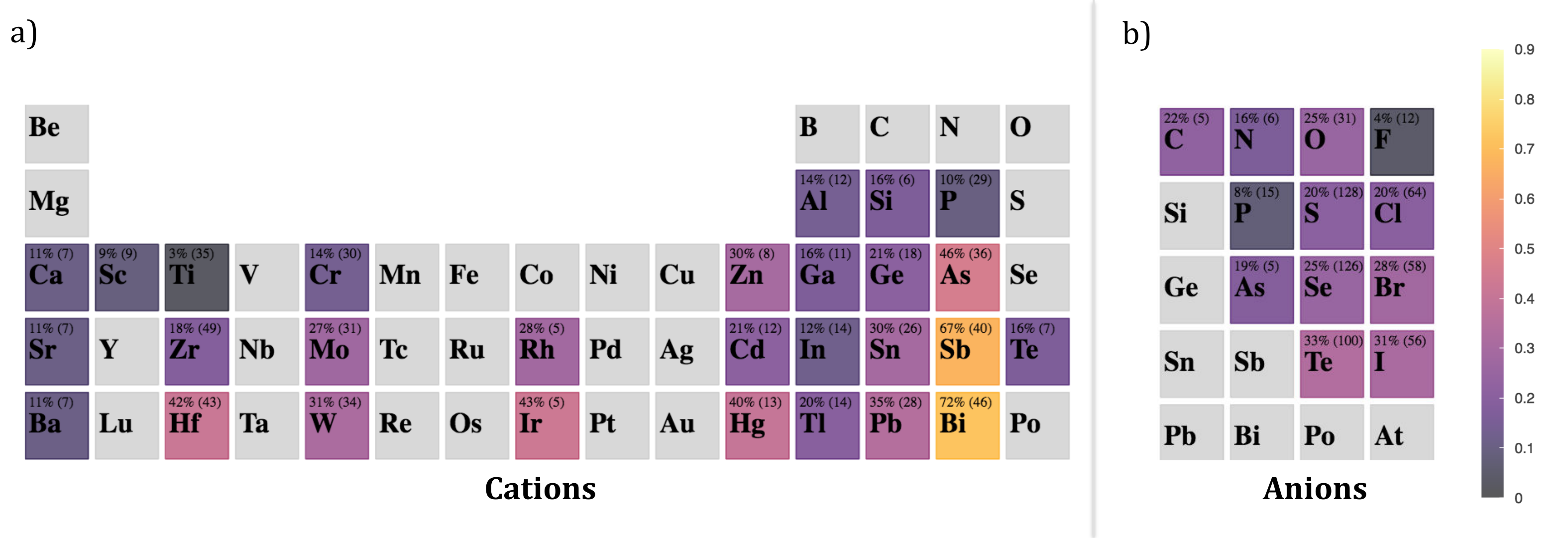}
\caption{Bayesian inference heatmap for Zeeman spin splitting greater than 100 meV given the presence of an element as a cation/anion in the material's composition. Each element cell also shows its posterior probability and its count (the number of materials with the element as cation/anion). The prior probability is 28\%.}
\label{fig:composition}
\end{figure}

\begin{table}[ht]
\centering
\begin{tabular}{|l|l|l|l|l|}
\hline
Target property & Bottom limit & Structural cluster & Cations & Anions \\
\hline
Zeeman spin splitting & 100 meV & \begin{tabular}[c]{l}Buckled Honeycomb AB\\T-TMDC \ce{MXY} Janus\\H-TMDC\\H-TMDC \ce{MXY} Janus\end{tabular} & \begin{tabular}[c]{l}Bi, Sb, As, \\ Ir, \\Hf\end{tabular} & \begin{tabular}[c]{l}I, \\Te\end{tabular} \\
\hline
Rashba/Dresselhaus spin splitting & 100 meV &\begin{tabular}[c]{l}T-TMDC \ce{MXY} Janus\\Buckled Hexagonal AB\\H-TMDC alloys\end{tabular} & \begin{tabular}[c]{l}Bi, Sb, As, \\W, Mo, Cr\end{tabular} & \begin{tabular}[c]{l}Te, Se\end{tabular} \\
\hline
Rashba parameter & 1.0 eV.Å &\begin{tabular}[c]{l}T-TMDC \ce{MXY} Janus\\H-TMDC alloys\end{tabular} & \begin{tabular}[c]{l}Bi, Sb, \\W, Mo, Cr\end{tabular} & \begin{tabular}[c]{l}Te, Se\end{tabular} \\
\hline
Band gap & 2.0 eV & \begin{tabular}[c]{l}\ce{AB3}-c19 \\Tetracoordinated \ce{MX_2} \\Buckled Hexagonal AB \\ H-TMDC\end{tabular} &\begin{tabular}[c]{l}Mg, Ca, Sr, Ba,\\Y,\\Cd, Zn\end{tabular}& \begin{tabular}[c]{l}F, Cl, Br, \\N\end{tabular} \\
\hline
\end{tabular}
\caption{Main results of Bayesian inference analysis targeting Rashba parameter, Zeeman spin splitting, and band gap. The bottom limit sets the value from which materials are classified as optimal. The following columns show the Crystal structures, cations, and anions which the Bayesian inference indicates as optimal characteristics for optimal target properties.}
\label{tab:results}
\end{table}

\begin{figure}[ht]
\centering
\includegraphics[width=\linewidth]{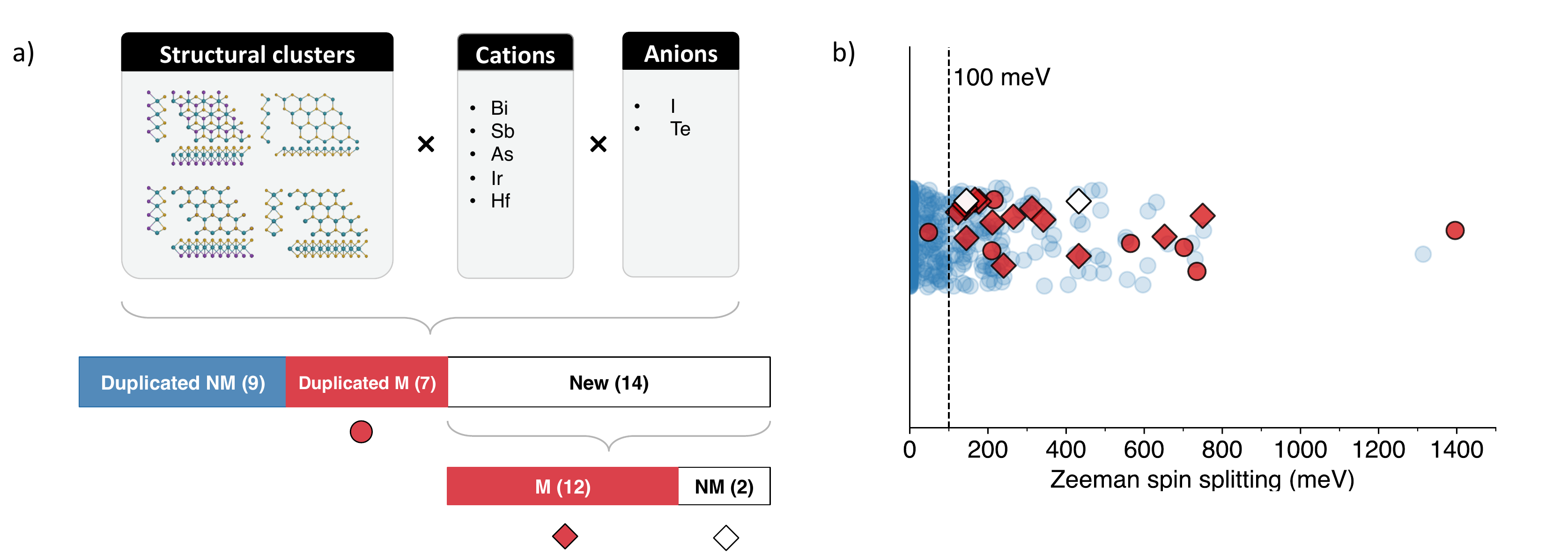}
\caption{a) Illustration of the combination of optimized structural clusters and optimized compositions, resulting in 30 combinations. 'M' (red) and 'NM' indicate materials that are metals and non-metals, respectively. Duplicated NM: combinations already presented in the screened dataset. Duplicated M (red circle): combinations presented in the non-screened dataset. New (diamonds): new materials generated by optimizing criteria; b) Strip plot of the Zeeman spin splitting. Blue circles correspond to all Zeeman spin splittings for all materials in the dataset (materials that do not have ZSS are represented by a Zeeman spin splitting of zero). White diamonds: new materials with band gap > 0.}
\label{fig:dist_zss}
\end{figure}

% from which nine were already in the screened dataset of 436 non-metal materials (Duplicated NM), seven were in the general C2DB database as metals (Duplicated M), and 14 are new combinations (New).

\begin{table}[]
\centering
\begin{tabular}{|l|l|l|l|l|}
\hline
Composition                 & Crystal structure & Zeeman spin splitting (meV) & Accessibility (meV) & Ehull (meV) \\ \hline
\ce{WTe2}   & H-TMDC            & 485.2                       & 0.0                 & 26.3        \\ \hline
\ce{WSe2}   & H-TMDC            & 466.1                       & 0.0                 & 0.0         \\ \hline
\ce{WSSe}   & H-TMDC MXY Janus            & 445.1                       & 0.0                 & 10.2        \\ \hline
\ce{WS2}    & H-TMDC            & 430.0                       & 0.0                 & 0.0         \\ \hline
\ce{MoTe2}  & H-TMDC            & 215.1                       & 0.0                 & 0.0         \\ \hline
\ce{MoSeTe} & H-TMDC MXY Janus            & 199.8                       & 0.0                 & 25.0        \\ \hline
\ce{MoSe2}  & H-TMDC            & 184.6                       & 0.0                 & 0.0         \\ \hline
\ce{MoSSe}  & H-TMDC MXY Janus            & 168.5                       & 0.0                 & 9.3         \\ \hline
\ce{MoS2}   & H-TMDC            & 148.0                       & 0.0                 & 0.0         \\ \hline
\ce{ZrI2}   & H-TMDC            & 121.8                       & 0.0                 & 27.0        \\ \hline
\ce{Al2Te2} & InSe            & 118.0                       & 55.4                & 0.0         \\ \hline
\ce{CrSe2}  & H-TMDC            & 90.3                        & 0.0                 & 0.0         \\ \hline
\ce{CrSSe}  & H-TMDC MXY Janus  & 81.5                        & 0.0                 & 10.4        \\ \hline
\ce{ZrBr2}  & H-TMDC            & 75.8                        & 0.0                 & 0.0         \\ \hline
\ce{CrS2}   & H-TMDC            & 68.3                        & 0.0                 & 0.0         \\ \hline
\ce{ZrClBr} & H-TMDC MXY Janus  & 59.4                        & 0.0                 & 9.8         \\ \hline
\ce{Al2Se2} & InSe             & 56.3                        & 62.8                & 0.0         \\ \hline
\ce{TiBr2}  & H-TMDC            & 55.9                        & 0.0                 & 0.0         \\ \hline
\ce{TiClBr} & H-TMDC MXY Janus  & 45.4                        & 0.0                 & 15.3        \\ \hline
\ce{ZrCl2}  & H-TMDC            & 40.8                        & 0.0                 & 0.0         \\ \hline
\ce{TiCl2}  & H-TMDC            & 31.7                        & 0.0                 & 0.8         \\ \hline
\end{tabular}
\caption{{\it Best-of-a-class} ZSS materials according to Zeeman spin splitting, accessibility (how far from the band edge the splitting occurs) and ehull (energy above hull) values.}
\label{tab:best_zss}
\end{table}

% \begin{figure}[ht]
% \centering
% \includegraphics[width=\linewidth]{stream}
% \caption{Legend (350 words max). Example legend text.}
% \label{fig:example}
% \end{figure}

% \begin{table}[ht]
% \centering
% \begin{tabular}{|l|l|l|}
% \hline
% Condition & n & p \\
% \hline
% A & 5 & 0.1 \\
% \hline
% B & 10 & 0.01 \\
% \hline
% \end{tabular}
% \caption{\label{tab:example}Legend (350 words max). Example legend text.}
% \end{table}

%\begin{figure}[h!]
%  \centering
%  \includegraphics[width=0.6\linewidth]{anticrossing_scatter.png}
%  \captionof{figure}{Distribution of the 124 compounds with Rashba or Dresselhaus SS prototypes identified in this work, according to its Rashba parameter $\alpha_R$ and its bandgap. Filled blue (unfilled black) dots correspond to compounds whose SS were identified to have (do not have) the presence of anti-crossing among valence and conduction bands.}
%  \label{fig:ac_analysis}
%\end{figure}
%
%\begin{figure}[h!]
%    \centering
%    \includegraphics[width=0.7\linewidth]{anti-crossing_01.png}
%    \caption{Representation of a compound where anti-crossing among deeper bands may have significant consequences for the SS observed at the band edges, for the case of entry with C2DB uid AsIS-b13beafa16aa. The inset represents the $p_z$ orbital character from the \ce{S} site.}
%    \label{fig:anti-crossing_01}
%\end{figure}

\begin{figure}[h!]
    \centering
    \includegraphics[width=\linewidth]{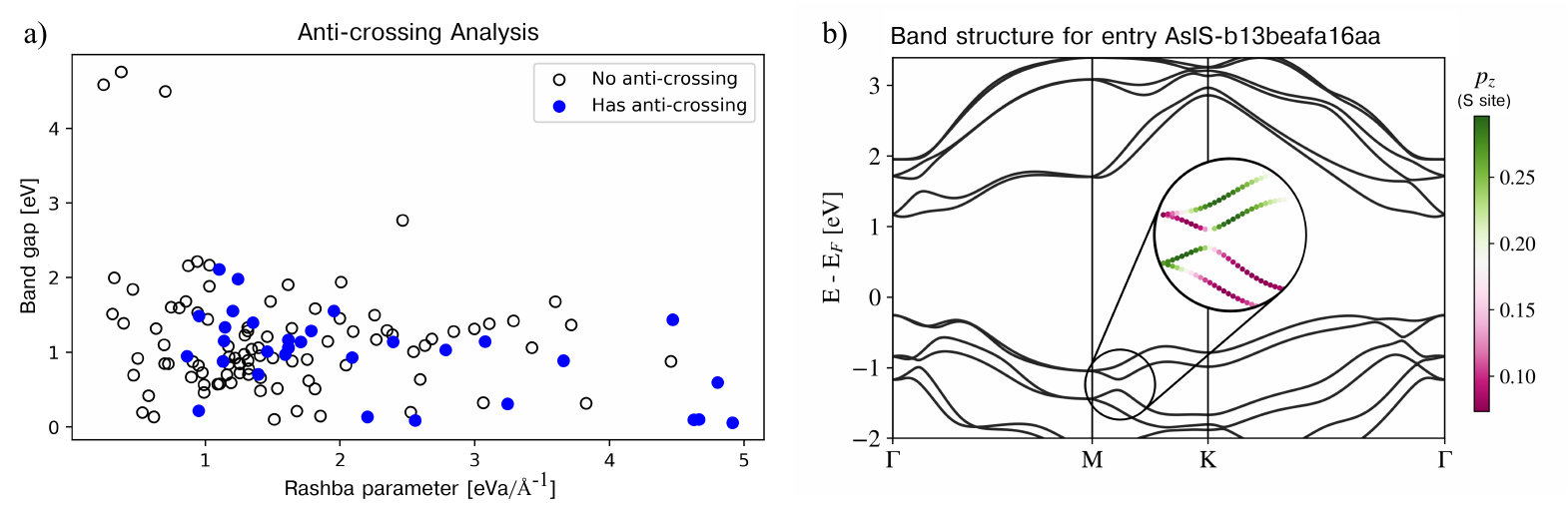}
    \caption{Presence of anti-crossing bands. a) Distribution of the 124 compounds with Rashba or Dresselhaus SS prototypes identified in this work, according to its Rashba parameter $\alpha_R$ and its band gap. Filled blue (unfilled black) dots correspond to compounds whose SS were identified to have (do not have) the presence of anti-crossing among valence and conduction bands. b) Representation of a compound where anti-crossing among deeper bands may have significant consequences for the SS observed at the band edges, for the case of entry with C2DB uid AsIS-b13beafa16aa. The inset represents the $p_z$ orbital character from the \ce{S} site.}
    \label{fig:anticrossing_figure}
\end{figure}

\clearpage

\includepdf[pages=-]{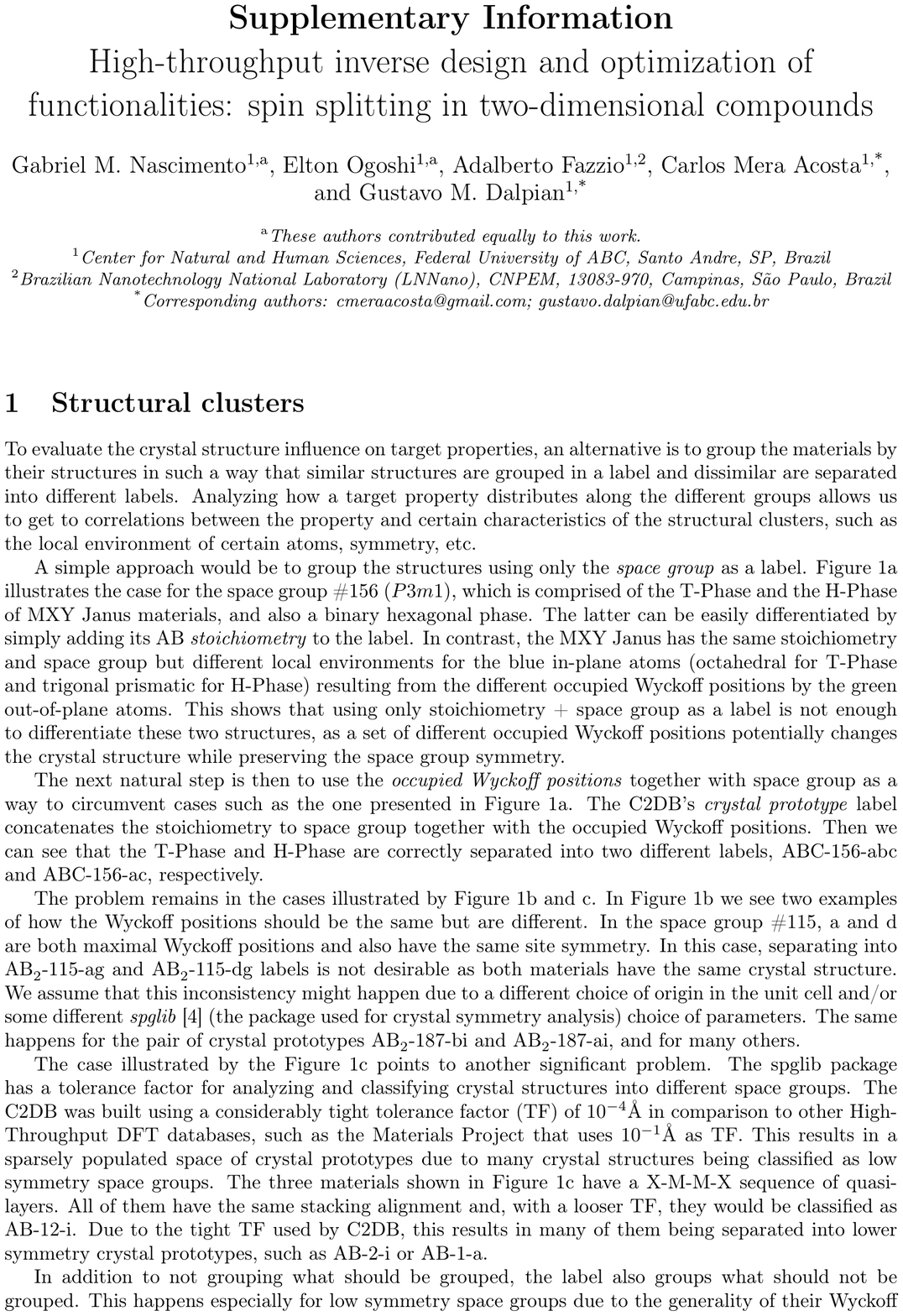}
\includepdf[pages=-]{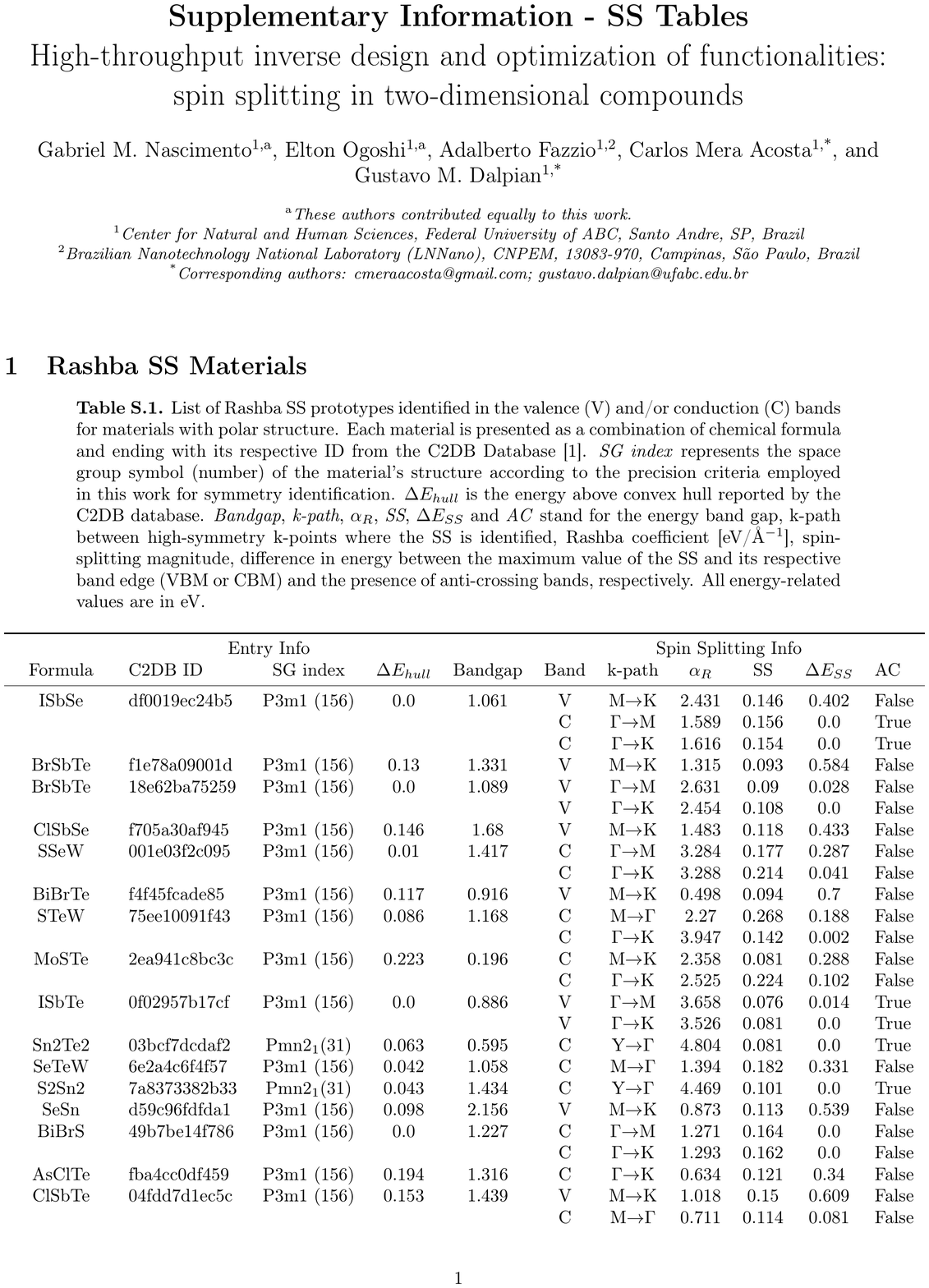}
\includepdf[pages=-]{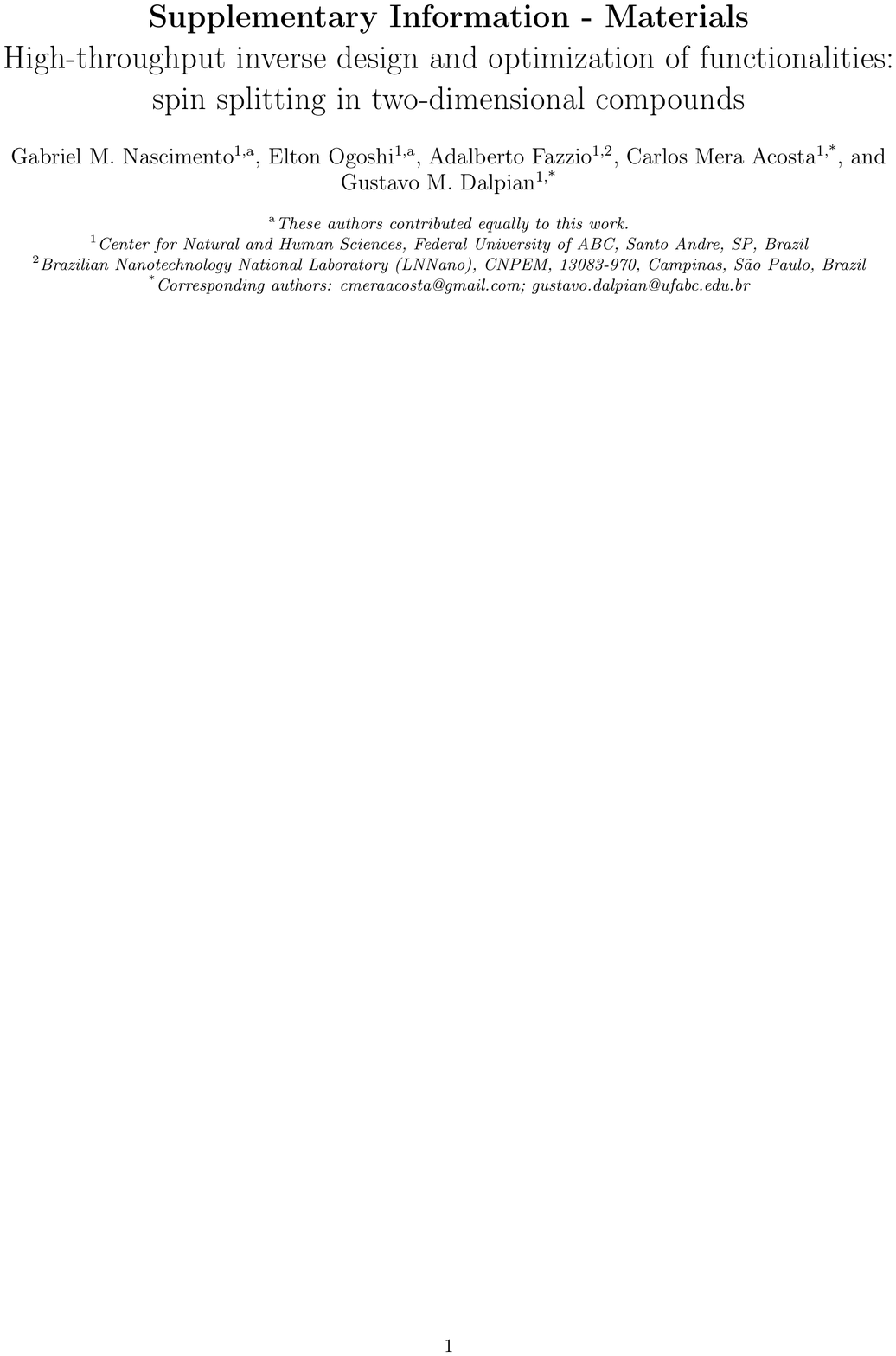}

\end{document}